\documentclass[aps,pra,superscriptaddress,12pt,tightenlines,nofootinbib]{revtex4}
\usepackage{amsmath,amsthm,graphicx,amssymb,verbatim,listings}
\usepackage[T1]{fontenc}     % needed for the guillemets
\usepackage{lmodern}         % needed to fix suboptimal font rendering
\usepackage[ pdftex, plainpages = false, pdfpagelabels,
                 pdfpagelayout = useoutlines,
                 bookmarks,
                 bookmarksopen = true,
                 bookmarksnumbered = true,
                 breaklinks = true,
                 linktocpage=all,
                 pagebackref=false,
                 colorlinks = true,
                 linkcolor = BrickRed,
                 urlcolor  = blue,
                 citecolor = BrickRed,
                 anchorcolor = green,
                 hyperindex = true,
                 hyperfigures
                 ]{hyperref}
\usepackage[usenames, dvipsnames]{xcolor}
\usepackage{xifthen} % provides \isempty test

\newcommand{\ket}[1]{{\ensuremath{\left| #1 \right\rangle}}}

\newcommand{\arxiv}[2][]{\ifthenelse{\isempty{#1}}{\href{http://arxiv.org/abs/#2}{{\tt arXiv:\allowbreak{}#2}}} {\href{http://arxiv.org/abs/#2}{{\tt arXiv:\allowbreak{}#2 [#1]}}}}

\newcommand{\booktitle}{\textsl}
\newcommand{\hrefdoi}[2]{\href{https://dx.doi.org/#1}{#2}}

\begin{document}

\title{Is Relational Quantum Mechanics about Facts? If So, Whose? A Reply to Di Biagio and Rovelli's Comment on Brukner and Pienaar}

\author{Blake C.\ Stacey}
\affiliation{Physics Department, University of Massachusetts Boston}

\date{\today}

\begin{abstract}
  Brukner and Pienaar have critiqued the Relational Quantum Mechanics
  of Rovelli, and together with Di Biagio, the latter has replied. I
  point out a few places where, in my view, that reply needs
  clarification.
\end{abstract}

\maketitle

Relational Quantum Mechanics is an interpretation of quantum theory
championed by Rovelli~\cite{Rovelli:1996, Rovelli:2021}. Recently,
Brukner and Pienaar separately wrote critiques of
RQM~\cite{Brukner:2021, Pienaar:2021, Pienaar:2021b}, to which Di
Biagio and Rovelli have written a reply~\cite{DiBiagio:2021}. This
exchange is a welcome step towards clarity in the murky business of
quantum interpretations, but areas of opacity persist.

In order to ground the discussion, Pienaar provides six claims, or
premises, intended to define RQM. Di Biagio and Rovelli take exception
to this list in various ways. For the present purposes, we can start
with Pienaar's fifth premise (emphasis in the original):
\begin{quote}
\textbf{Any physical correlation is a measurement.} Suppose an
observer measures a pair of systems and thereby assigns them a joint
state which exhibits perfect correlations between some physical
variables. Then the two systems have measured each other (entered into
a measurement interaction) relative to the observer, and the physical
variables play the roles of the `pointer variable' and `measured
variable' of the systems.
\end{quote}
Here is what Di Biagio and Rovelli call a ``proper reformulation'' of
Pienaar's fifth premise (emphasis again as in the original):
\begin{quote} 
\textbf{An interaction between two systems results in a correlation
  within the interactions between these two systems and a third one.}
With respect to a third system $\mathcal{W}$, the interaction between
the two systems $\mathcal{S}$ and $\mathcal{F}$ is described by a
unitary evolution that potentially entangles the quantum states of
$\mathcal{S}$ and~$\mathcal{F}$.
\end{quote}
The letters $\mathcal{W}$ and $\mathcal{F}$ refer to Wigner and his
Friend respectively. This version of the premise prompts at least
three concerns. First, Di Biagio and Rovelli criticize Brukner and
Pienaar for putting too much emphasis on ``states'' rather than
``facts'', but here, they spell out the meaning of the ``proper
reformulation'' in the language of states and unitaries. Consequently,
it is hard to tell what difference of \emph{substance} there is
between Pienaar's version and theirs. This is particularly
troublesome, because they call this the ``main problem with Pienaar's
account'', and so identifying that difference of substance is
particularly important. Second, in turn, this lapse back into the
language of states and unitaries lends weight to the concern that,
while the RQM literature claims to de-ontologize quantum states, in
practice it turns around and reifies them again. I have written at
greater length about this point elsewhere~\cite{Stacey:2021}; it is a
species of a more general trouble with attempts to interpret quantum
theory in informational terms that only go halfway~\cite{Fuchs:2020,
  Fuchs:2020b}. Third, in the bolded sentence, the word
``interaction'' is being used in two different senses. This introduces
an ambiguity that is not present in Pienaar's version. Given systems
$\mathcal{W}$, $\mathcal{F}$ and $\mathcal{S}$, there are
``interactions'' between $\mathcal{F}$ and $\mathcal{S}$ to which
$\mathcal{W}$ is a bystander, and then there are the possible
``interactions'' between $\mathcal{W}$ and $\mathcal{F}$ or
$\mathcal{S}$ or the $\mathcal{F}$--$\mathcal{S}$ pair. The latter
class of ``interactions'' constitute ``events'' relative to
$\mathcal{W}$, but the former do not. The ``interactions'' for which
$\mathcal{W}$ is a bystander can be described by a Hamiltonian,
whereas those involving $\mathcal{W}$ itself are more like Wheeler's
\emph{elementary acts of fact creation}~\cite{Wheeler:1982}. (Perhaps,
following Barad, those in which $\mathcal{W}$ itself participates
could be called ``intra-actions'' for contrast~\cite{Barad:2007}.)

We can underline this by considering a passage from their
introduction, where Di Biagio and Rovelli write,
\begin{quote}
The quantum state of a composite system relative to an external system
is not an account or record of relative events \emph{between the
  subsystems} of the composite system. It is only a mathematical tool
useful for predicting probabilities of events \emph{relative to the
  external system.}
\end{quote}
If we hold fast to this, then the term ``interaction'' in their
version of the fifth premise is necessarily overloaded. So, in their
``proper reformulation'', we have one statement that is ambiguous, and
then a nominal clarification of it in the language of states and
unitaries, which is the language we are told we should
\emph{de}-emphasize in favor of ``events'' or ``facts''.

Likewise, it is difficult to distinguish Pienaar's sixth premise from
Di Biagio and Rovelli's replacement for it. Here is Pienaar's:
\begin{quote}
In the Wigner's friend scenario as outlined above, if $\mathcal{W}$
measures $\mathcal{F}$ to `check the reading' of a pointer variable
(i.e.\ by measuring $\mathcal{F}$ in the appropriate `pointer basis'),
the value he finds is necessarily equal to the value that
$\mathcal{F}$ recorded in her account of her earlier measurement of
$\mathcal{S}$.
\end{quote}
And here is the replacement:
\begin{quote}
In the Wigner's friend scenario, if $\mathcal{W}$ measures
$\mathcal{S}$ on the same basis on which $\mathcal{F}$ did, then
appropriately interacts with $\mathcal{F}$ to `check the reading' of a
pointer variable (i.e.\ by measuring $\mathcal{F}$ in the appropriate
`pointer basis'), the two values found are in agreement.
\end{quote}
Di Biagio and Rovelli say that Pienaar's version is ``either wrong
\ldots\ or a tautology'', but their replacement suffers the same
problem. The phrase ``if $\mathcal{W}$ measures $\mathcal{S}$ on the
same basis on which $\mathcal{F}$ did'' is evidently problematic,
since it presumes that the choice of basis made by $\mathcal{F}$ is a
publicly available fact, viewable from everywhere and nowhere. Di
Biagio and Rovelli's version avoids saying that ``the value that
$\mathcal{F}$ recorded'' is a fact for all, but as long as the choice
of basis is still treated that way, the claim is still contradictory,
or at best, ambiguous~\cite{DeBrota:2018}.

A later passage appears at first glance to address this:
\begin{quote}
So how do we know which of $\mathcal{S}$'s variables became definite
relative to $\mathcal{F}$? We do not, if we only know the state
$\ket{\Psi}_{\mathcal{SF}}$.
\end{quote}
So far, this would suggest that indeed, we should not take
$\mathcal{F}$'s choice of basis as a fact for all. But then:
\begin{quote}
More information can be obtained from the dynamics of the system.  The
state (1) for instance may arise as a result of an interaction between
$\mathcal{S}$ and $\mathcal{F}$ in which the evolution of
$\mathcal{F}$ depends on the value of the variable $X$ of
$\mathcal{S}$. For example, the interaction Hamiltonian can depends
[sic] on this variable. From the perspective of $\mathcal{F}$, this
interaction leads to the actualization of the variable $X$ of
$\mathcal{S}$. But the same state relative to $\mathcal{W}$ could
arise via an interaction Hamiltonian that depends on the variable $Y$,
and then it is this variable that actualizes relative to
$\mathcal{F}$.
\end{quote}
Is the interaction Hamiltonian not relative? Are unitaries and their
generators somehow absolute? Di Biagio and Rovelli's ``proper
reformulation'' of the fifth premise declares that the unitary
evolution describing the time evolution of
$\ket{\Psi}_{\mathcal{SF}}$ is with respect to $\mathcal{W}$, just
like the state vector is. How, then, is the generator of that unitary
at all binding upon what might actualize with respect to
$\mathcal{F}$? Conversely, how does an interaction Hamiltonian that is
true with respect to $\mathcal{F}$ evolve the state vector that is
true with respect to $\mathcal{W}$?

Much as their fifth premise backs away from a radical position
regarding quantum states, their elaboration upon their sixth premise
backs away from the relativity of physical facts that is supposed to
characterize the interpretation. Given this resolute retreat from
audacity, it seems odd to accuse Pienaar of disliking RQM for ``being
more radical than what he hoped for'', all the more so given his known
QBist sympathies~\cite{Pienaar:2020, Pienaar:2020b, DeBrota:2021}.

\bigskip

This note is based on remarks originally made at
\href{https://scirate.com/arxiv/2110.03610}{SciRate}.

\end{document}